\documentclass[conference]{IEEEtran}
\IEEEoverridecommandlockouts
\usepackage{cite}
\usepackage{amsmath,amssymb,amsfonts}
\usepackage{algorithmic}
\usepackage{graphicx}
\usepackage{textcomp}
\usepackage{xcolor}
\usepackage{booktabs}
\usepackage{multirow}
\usepackage{color}
\usepackage{adjustbox}
\usepackage[ruled,vlined]{algorithm2e}
\usepackage{hyperref}

\def\BibTeX{{\rm B\kern-.05em{\sc i\kern-.025em b}\kern-.08em
    T\kern-.1667em\lower.7ex\hbox{E}\kern-.125emX}}
\begin{document}


\title{Causal Inference in Finance: An Expertise-Driven Model for Instrument Variables Identification and Interpretation}

\author{\IEEEauthorblockN{1\textsuperscript{st} Ying Chen}
\IEEEauthorblockA{\textit{Tokyo Institute of }\\
\textit{Technology}\\
Tokyo, Japan \\
chen.y.cc3c@m.isct.ac.jp}
\and
\IEEEauthorblockN{2\textsuperscript{nd} Ziwei Xu}
\IEEEauthorblockA{\textit{National Institute of Advanced }\\
\textit{Industrial Science and Technology}\\
Tokyo, Japan \\
xu.ziwei@aist.go.jp}
\and
\IEEEauthorblockN{3\textsuperscript{rd} Kotaro Inoue}
\IEEEauthorblockA{\textit{Tokyo Institute of }\\
\textit{Technology}\\
Tokyo, Japan \\
inoue.k.aq@m.titech.ac.jp}
\and
\IEEEauthorblockN{4\textsuperscript{th} Ryutaro Ichise}
\IEEEauthorblockA{\textit{Tokyo Institute of }\\
\textit{Technology}\\
Tokyo, Japan \\
ichise@iee.e.titech.ac.jp}
}

\maketitle

\begin{abstract}
Instrumental Variable (IV) provides a source of treatment randomization that is conditionally independent of the outcomes, responding to the challenges of counterfactual and confounding biases. In finance, IV construction typically relies on pre-designed synthetic IVs, with effectiveness measured by specific algorithms. This classic paradigm cannot be generalized to address broader issues that require more and specific IVs. Therefore, we propose an expertise-driven model (ETE-FinCa) to optimize the source of expertise, instantiate IVs by the expertise concept, and interpret the cause-effect relationship by integrating concept with real economic data. The results show that the feature selection based on causal knowledge graphs improves the classification performance than others, with up to a 11.7\% increase in accuracy and a 23.0\% increase in F1-score. Furthermore, the high-quality IVs we defined can identify causal relationships between the \emph{treatment} and \emph{outcome} variables in the Two-Stage Least Squares Regression model with statistical significance.

\end{abstract}


\section{Introduction}

\begin{figure}[htbp]
\centerline{\includegraphics[width=0.5\textwidth]{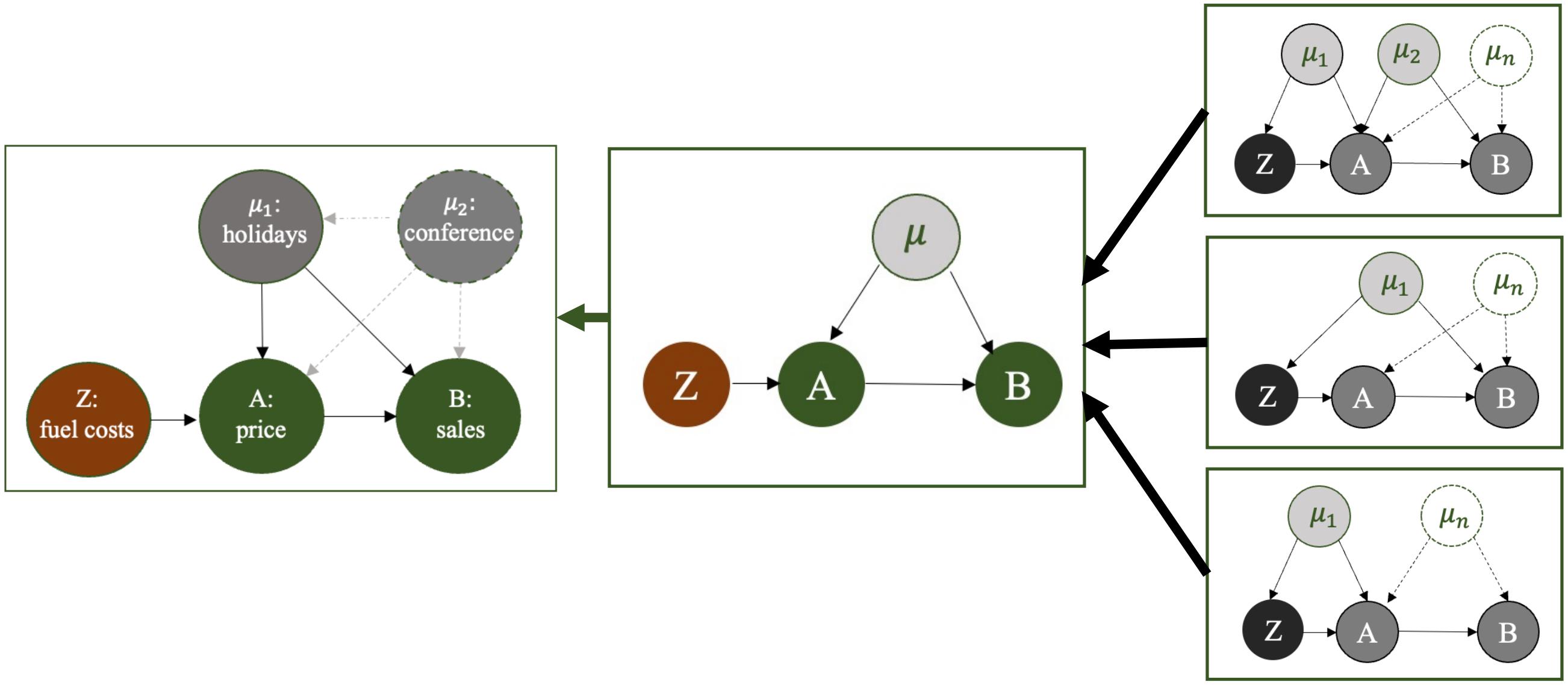}}
\caption{The causal graphs with instrument variable (Z) specification.}
\label{fig1}
\end{figure}

The instrumental variable (IV) approach provides a source of treatment randomization that is conditionally independent of the outcome to estimate the counterfactual effect using observational data. In Figure~\ref{fig1}, we take the airline ticket demand scenario to explain the relationships among \emph{instrumental variable}(\( Z \)), \emph{treatment}(\( A \)), \emph{outcome}(\( B \)) and \emph{other observed/ unobserved variables}(\( \mu_{n}\)). Typically, lower ticket prices lead to higher sales, yet high sales can also occur during holidays ($\mu_{1}$) despite high ticket prices. The arrow from A to B indicates that price causes sales, with observed holiday ($\mu_{1}$) affecting both. Other unobserved variables like conferences ($\mu_{2}$) also impact A and B, adding complexity to their direct causality. However, by including an IV (Z) that solely impacts A but not B directly, we can clarify the causality between A and B. This aligns with the principle that fuel costs (Z) affect ticket prices (A), which then influences sales (B). This clear causal pathway, enabled by including the IV, excludes the possibility of other variables directly affecting outcomes, providing valuable insights for financial experts conducting quantitative research.

In finance, IVs construction typically relies on pre-designed synthetic IVs, with effectiveness measured by specific algorithms. This classic paradigm cannot be generalized to address broader issues that require more and specific IVs. For instance, in the traditional Two-Stage Least Squares (2SLS) method, which is widely accepted in empirical economic research, their coefficients can be simply interpreted under strict theoretical assumptions as follows: ceteris paribus, a one-unit increase in \emph{treatment} (\( A_i \)) will literally cause a \( \beta_i \) change in the \emph{outcome} (\( B_i \)). In practical applications, the strict assumptions of correct specification and exogeneity can only be relaxed with the knowledge of economists. Therefore, we propose to identify IVs directly from textual expertise, instantiate IVs by their expertise concept, and interpret the cause-effect relationship by the significant IVs calculated from real economic data with 2SLS regression model. We abbreviate our method, \textit{\textbf{E}xper\textbf{T}ise-driven mod\textbf{E}l for \textbf{Fin}ancial \textbf{Ca}usal variables identification and interpretation} as \textbf{ETE-FinCa}, which will be used throughout the following sections.


Furthermore, \textbf{ETE-FinCa} has shown significant effectiveness in two key tasks: 

\textit{\textbf{(1) Which dataset possesses adequate expertise for IV identification tasks?}} We introduce various corpus sources to retrieve important features. Using the predefined IV classification task, we assess the classification models on these features to identify the corpus with optimal expertise. The results show that the feature selection based on our approach improves the classification performance than others, with up to a 11.7\% increase in accuracy and a 23.0\% increase in F1-score.  

\textbf{\textit{(2) How are IVs interpreted for causal prediction? }} The causal relationships that appear in all subgraphs are considered common sense. We design a task to focus on those causal variables that are only mentioned in certain graphs due to divergent standpoint. These insufficiently studied standpoint-based causality may face more complex confounding biases. We develop an interpretation module to curate 19,678 causal structures that are consistent with the IV pattern from the causal knowledge graph, identify high-quality IVs that exclusively exist in certain subgraphs and interpret the directional causal relationships extracted from different standpoints with real economic data through 2SLS Regression. We generalize the standpoint-based causality and demonstrate the research potential of these specific expertise in financial domain.


\section{Related Work}

\subsection{Causal Inference and Instrument Variables (IV)}


Recent work aim to understand the impact of confounders through observational data without performing randomized experiments. This framework includes the potential outcome framework \cite{b2} and the structural causal model (SCM) \cite{b4}. The former is also known as the Rubin Causal Model, which aims to estimate potential outcomes and subsequently calculate the treatment effect. Meanwhile, SCM approach describes the causal mechanisms of a system where variables and their causal relationships are modeled using a set of simultaneous structural equations or causal graphs. Confounders lead to incorrect causal relationships when estimating the Average Treatment Effect (ATE) of interventions. Thus, mitigating confounding bias is critical in causal inference. Prior research has proposed several methods to mitigate selection bias and simulate the true distribution of the target group, such as sample reweighting \cite{b5}, stratification \cite{b1}, matching \cite{b11}, tree-based methods \cite{b15}, representation\cite{b16}, and multitask methods \cite{b12}. Another effective approach is to train a foundational estimator of potential outcomes using observational data and then correct for estimation bias caused by selection bias \cite{b17}. Similar to \textbf{ETE-FinCa}, previous approaches apply intervention effects conditioned on confounders and perform weighted averaging based on their distribution \cite{b9}. 

Instrument variable \( Z \) only influences outcomes \( B \) via variables \( A \), which allows for the identification of directional causal relationships even in the presence of confounding biases. The classic 2SLS regression requires the researcher to have a strong prior understanding, and this method is constrained by computational complexity. Advanced machine learning methods have demonstrated the power of learning potential representations of complex feature spaces. Recent remarkable works include Deep-IV \cite{b6}, Kernel-IV \cite{b13}, and Auto-IV \cite{b7}. DeepIV trains a network to estimate the conditional distribution of treatment variable given the instruments and covariance, integrating this into a causal inference network to estimate causal effects. Kernel-IV relaxes linearity assumptions by modeling IV patterns with nonlinear functions in Reproducing Kernel Hilbert Spaces (RKHS). AutoIV generates confounder representations from observational data and inputs these IV candidates into an adversarial game network with mutual information maximization and minimization constraints until the IV candidates meet relevance and exclusion conditions. The lack of interpretability limits the practical application of these methods.

\subsection{Interpretable Approaches and Causal Knowledge Graph}

To address these issues, recent work focus on applying the invariant causal relationships from observable data to establish models that provide stable and interpretable predictions. These methods typically adhere to the unconfoundedness assumption, e.g., propensity score\cite{b21}, covariate balance\cite{b19}, back-door criteria\cite{b4}, and representation learning\cite{b20}. However, confounding bias is inevitable in practical problem settings. It is necessary to find a method to comprehensively display the interconnected effect of all potential and observable variables. \textbf{ETE-FinCa} maps the causal relationships between financial concepts and extracts causal chains that fit the IV pattern, enabling researchers or decision-makers to utilize the framework to understand the useful instrumental variables and their connections with other observed variables, thereby enhancing the interpretability. Causal relations have been explored in many open source knowledge bases, such as WikiData\cite{b3} and ConceptNet\cite{b14}. Recently, causality-dedicated knowledge graphs have been generated by many works, including CausalNet\cite{b22}, Cause Effect Graph\cite{b23}, CauseNet\cite{b18}, and ATOMIC\cite{b8}. The entire logic of thinking, including potential confounding factors, treatment, and outcomes, is crucial in causal inference. However, the causal relationships within these graphs lack interconnections between tuples, making it difficult to identify IV representations. FinCaKG\cite{b10} provides a causal knowledge graph based on cause-effect textual spans. The construction of this graph allows \textbf{ETE-FinCa} to use an end-to-end visualization framework for instrumental variable mining. Specifically, this method presents causal relationships through intuitive logical chains, simplifying the identification of instrumental variables, significantly enhancing the interpretability, and making the application of IV extracted from \textbf{ETE-FinCa} in economic empirical research possible.

\begin{figure*}[htbp]
\centerline{\includegraphics[width=\textwidth]{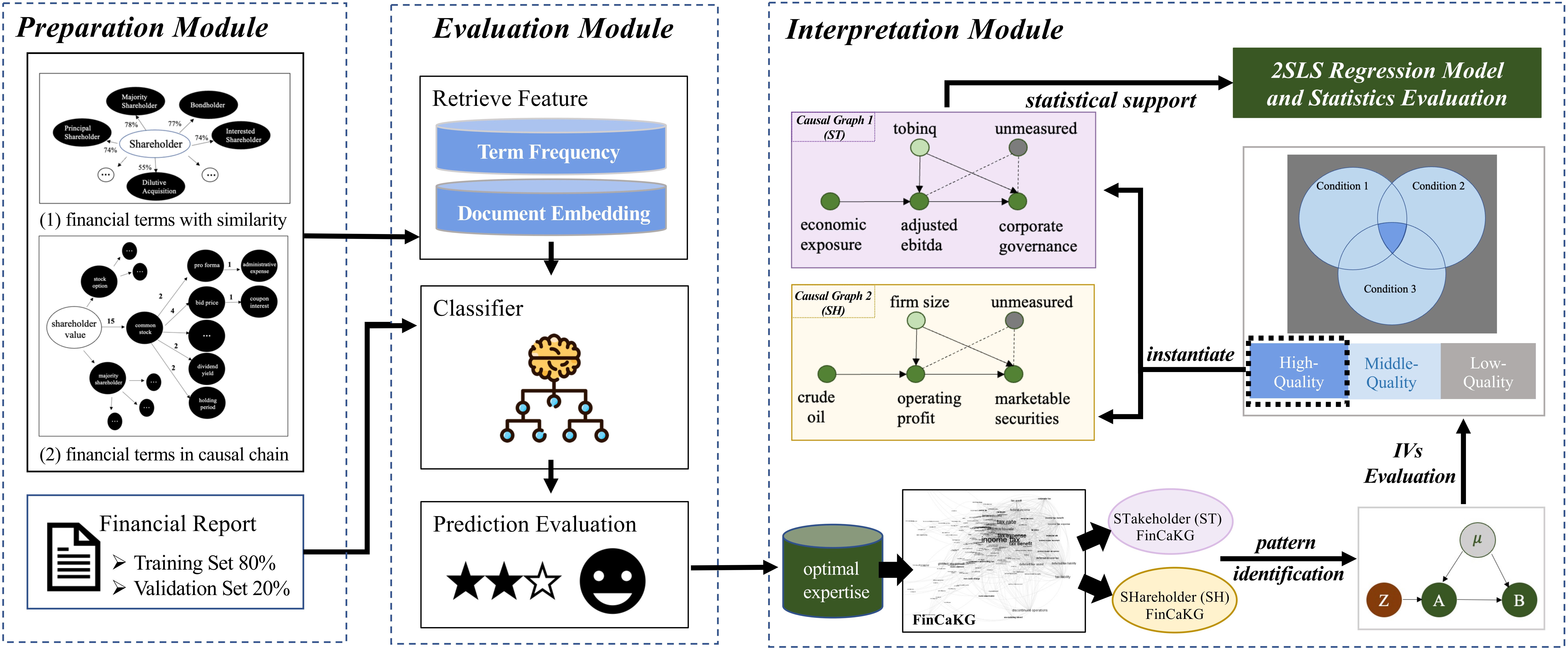}}
\caption{The workflow of \textbf{ETE-FinCa} model for causal vairables identification and interpretation.}
\label{fig2}
\end{figure*} 

\section{Methodology}

In this section, we outline the procedures from expertise selection to causal variables identification and interpretation of the \textbf{ETE-FinCa} model. As shown in Figure~\ref{fig2}, our method consists of three modules: Preparation Module, Evaluation Module, and Interpretation Module.

\begin{algorithm}
\caption{IV identification based on DFS}
\label{alg1}
\KwIn{Graph G, all\_NPs\_unique (V), tuple\_nodes\_sameid}
\KwOut{Set of (Z, A, B) triples K}
Define R(x, y): the distance between x and y in Graph is less than or equal to 3 hops\;
Initialize empty result set K\;
\For{each z in V}{
    $A_z = \{a \in V | R(z, a)\}$\;
    \For{each a in $A_z$}{
    
        $B_{temp}$ = $\{b \in V | R(a, b)\} \setminus a$\;
        
        \For{each b in $B_{temp}$}{
        $C = \{c \in V | R(b, c)\}$\;
        \If{z $\in$ C}{
            Remove $b$ from $B_{temp}$;
        }
    }
        $B_z = \{b \in B_{temp} | \neg R(z, b)\}$\;
    }
   
    Add triple (z, a, b) to result set K\;
}
\Return{K}
\end{algorithm}

\subsection{Preparation Module}

The corpus and labeled training sets are placed in the preparation module. we process financial vocabulary obtained from Investopedia\footnote{\url{https://www.investopedia.com/financial-term-dictionary-4769738}, access date:2024/03/28} that provides expertise related to finance, investing, and economic concepts. Regarding the first corpus, namely the cosine similarity-based corpus $C_{sim}$ (see Picture (1) in Figure~\ref{fig2}.), which involved computing cosine similarity by pipeline of spaCy between each concept in the vocabulary list and `shareholder'. The second corpus $C_{FinCaKG}$ (see Picture(2) in Figure 2) is based on a financial causal knowledge graph. We extract the terms from causal chains that span multiple hops to describe a logical pathway. The labelled annual reports from companies that claim to be either ``stakeholder-maximizing'' or ``shareholder-maximizing'' are compiled into a dataset, split into 80\% for the training set and 20\% for the validation set. 


\begin{table*}[t]
    \centering
    \caption{Classification Results}
    \label{tab1}
    \Huge
    \resizebox{\textwidth}{!}{\renewcommand{\arraystretch}{1.25}\begin{tabular}{lllccccc}
        \toprule
        \textbf{Corpus selection} & \textbf{Feature Style} & \textbf{Classifier Model} & \textbf{Num. of Feature} & \textbf{Accuracy} & \textbf{Precision} & \textbf{Recall} & \textbf{F1-score} \\
        \midrule
        \multirow{3}{*}{ $C_{sim}$} & \multirow{2}{*}{Term Frequency} & Random Forest & 166 & 0.726 & 0.707  & 0.426 & 0.524 \\
        & & XGBoost & 166 & 0.680 & 0.795  & 0.157 & 0.263  \\
        & Document Embedding (RoBERTa) & Random Forest & (max\_length,1024) & 0.713 & 0.752  & 0.336 & 0.464  \\
        \midrule
        \multirow{3}{*}{$C_{FinCaKG}$ (unweighted)} & \multirow{2}{*}{Term Frequency} & Random Forest & 602 & \underline{0.84} & \underline{0.872}  & \underline{0.655} & \underline{0.748} \\
        & & XGBoost & 602 & 0.744 & 0.779  & 0.411 & 0.538  \\
        & Document Embedding (RoBERTa) & Random Forest & (max\_length,1024) & 0.712 & 0.730  & 0.372 & 0.493  \\
        \midrule
        \multirow{3}{*}{$C_{FinCaKG}$ (weighted)} & \multirow{2}{*}{Term Frequency $+$ Weight} & Random Forest & 236 & \textbf{0.843} & \textbf{0.878}  & \textbf{0.660} & \textbf{0.754} \\
        & & XGBoost & 236 & 0.737 & 0.846  & 0.335 & 0.480  \\
        & Document Embedding (RoBERTa) & Random Forest & (max\_length,1024) & 0.718 & 0.733  & 0.379 & 0.500  \\
        \bottomrule
    \end{tabular}%
    }
    \begin{minipage}{\linewidth}
    \hfill \break
    \footnotesize{\textbf{Attention}: the first 1-3 rows are the baseline results. We mark the best-performing model with \textbf{bold markdowns} and the second-best model with \underline{underlines}}.
    \end{minipage}
\end{table*}

\subsection{Evaluation Module}
We classify two datasets, $C_{sim}$ and $C_{FinCaKG}$, beginning with feature retrieval. Suppose we have set of documents: $D = \{d_1, d_2, ..., d_n\}$, for each document $d_k \in D$, we remove the symbols and stopwords and obtain $d_k'$. For each document $d_k'$ and each term $t_i$ from $C_{sim}$ or $C_{FinCaKG}$, we have features:
\begin{equation}
    x_{i,k} = tf(t_i, d_k') \cdot weight(w_i)
\end{equation}

where $tf(t_i, d_k')$ represents the term frequency of $t_i$ in the document $d_k'$, $weight(w_i)$ is the cosine similarity in $C_{sim}$ and the edge weight of the relationship between two nodes in $C_{FinCaKG}$. Then we get $x_{i,k}$ = $\{x_1, x_2, ..., x_i\}$ the feature matrix for $d_k'$. The resulting feature matrix $X$ has dimensions $X \in \mathbb{R}^{n \times m}$, where $n$ is the number of nodes in each selected corpus and $m$ is the size of the corpus. The term frequency-based feature matrix is considered to be sparse. We also apply the transformer-based embedding method (RoBERTa) to learn the text representation because it handles long text efficiently and effectively without information loss. To conserve information as much as possible, we apply a length-weighted average, max pooling, direct concatenation, and a simple attention mechanism in the process of document embedding by using $RoBERTa$.




In the process of model evaluation, Random Forest and XGBoost are both popular choices for classification tasks in machine learning due to interpretability and robustness to overfitting. All model results are provided in terms of Accuracy (\%), Precision, Recall, and F1-score for evaluation.

\subsection{Interpretation Module}



In the interpretation module, we separate $C_{FinCaKG}$ to two subgraphs, ``shareholder-oriented standpoint''  ($C_{SH-FinCaKG}$) and ``stakeholder-oriented standpoint'' ($C_{ST-FinCaKG}$) and analyze the similarities and differences between them. 



Most causal chains lose causal meaning after three hops\cite{b24}; thus, we define associations within 3-hop as effective logical connections. Therefore, this task is defined as ``finding a Z within a 3-hop causal chain associated to A and not included in the causal chains associated to B within 3-hop.'' We execute a Depth-First Search (DFS) algorithm (see Algorithm~\ref{alg1}) to find patterns that meet the conditions of ``$A \perp B \mid Z$'' in FinCaKG. We first define a function \( R \) to search for all neighbor nodes within three hops of start node \( n \). We assume each node can potentially become an instrumental variable for other ``\textit{treatment-outcome}'' pairs. Therefore, we loop through all nodes as \( z \), use \( R(z, a) \) to find set \( A_z \) related to \( z \), then loop through the elements of set \( A \) and use ``\( R(a, b) \) and \(\neg R(z, b)\) '' to define set \( B \). Finally, we output the triple \(\{z, a, b\}\) to visualize the IVs.

In terms of IV evaluation, we categorize extracted IVs as ``\textit{high-quality (3 points)}'', ``\textit{middle-quality (1-2 points)}'', and ``\textit{low-quality (0 point)}'' from \textbf{ETE-FinCa}, according to the following scoring conditions: 
    \begin{itemize}
        \item {Z} is an edge node (+1 point);
        \item The weight between \textit{Z} and \textit{A}: $w_{z,a}$ $\geq 5.0$ (+1 point).
        \item The weight between \textit{A} and \textit{B}: $w_{a,b}$ $\geq 5.0$ (+1 point).
    \end{itemize}

Furthermore, we aim to identify high-quality IVs that exist only in certain subgraphs and run 2SLS regression model to interprets the directional effects and significance of ``\textit{treatment-outcome}'' variables and finally generalizes the specific expertise. The statistical results of 2SLS are calculated to interpret the validation of causality.


\section{Experiments and Results}


This section is dedicated to testing our proposed \textbf{ETE-FinCa} with diverse experimental configurations for classification and IV-identification tasks. Also, we apply an acceptable 2SLS regression model for the case study to provide insights for statistical supports in the financial domain.

\subsection{Data Preparation}

In the experiment, we collect 3,000 samples from globally listed companies that provided complete financial data for five consecutive years. The stock price, crude oil, and foreign exchange data were obtained from Bloomberg. The 2SLS regression model includes 14,099 observations. We also control for firm size and Tobin's Q as observed confounding variables to correct for sample bias. We set a threshold of 0.55 for \(C_{sim}\). A total of 2,436 financial concepts are included in the overall FinCaKG graph, with 1,890 nodes included in the ST-FinCaKG, and 1,566 nodes included in the SH-FinCaKG.







\subsection{Results}

\subsubsection{Classification}

Table~\ref{tab1} gives an overview of the results of our experiments using $C_{sim}$, $C_{FinCaKG}$(unweighted and weighted) to select the corpus for feature retrieval. $C_{FinCaKG}$ (unweighted) captures the most features (602) and significantly improves baseline performance. The feature matrix based on $C_{FinCaKG}$ (weighted) achieves best-performance compared to the baseline (see row 1 in $C_{sim}$), resulting in a 11.7\% increase in accuracy, a 17.1\% increase in precision, a 23.4\% increase in recall and a 23.0\% increase in F1 score (see row 1 in $C_{FinCaKG}$ (weighted)). In knowledge graphs, important information tends to be connected to multiple nodes and thus leads to higher edge weights. Compared to $C_{FinCaKG}$ (unweighted), $C_{FinCaKG}$ (weighted) improves the attention of model to 236 crucial expertise. Furthermore, the term frequency based on $C_{FinCaKG}$ (weighted) outperforms the document embedding approach (see row 3 in the $C_{sim}$) with an accuracy improvement of 13\%, a precision improvement of 12.6\%, a recall improvement of 32.4\%, and an F1 score improvement of 29\%. 
\begin{table}[t]
\caption{DFS Results}
\renewcommand{\arraystretch}{0.5}
\label{tab2}
\begin{tabular}{cccccc}
\toprule
\textbf{Dataset} & \textbf{Chain Pattern} & \textbf{Min.} & \textbf{Avg.± std.} & \textbf{Max.} & \textbf{Total} \\
\midrule
\multirow{3}{*}{All-FinCaKG} 
& Z & & & & 2,436 \\
\cmidrule{2-6}
& Z $\rightarrow$ A & 0 & 5±6 & 28 & 7,498 \\
\cmidrule{2-6}
& Z $\rightarrow$ A $\rightarrow$ B & 0 & 13±19 & 89 & 19,678 \\
\midrule
\multirow{3}{*}{ST-FinCaKG}
& Z & & & & 1,890 \\
\cmidrule{2-6}
& Z $\rightarrow$ A & 0 & 4±5 & 33 & 5,217 \\
\cmidrule{2-6}
& Z $\rightarrow$ A $\rightarrow$ B & 0 & 11±18 & 91 & 13,533 \\
\midrule
\multirow{3}{*}{SH-FinCaKG}
& Z & & & & 1,566 \\
\cmidrule{2-6}
& Z $\rightarrow$ A & 0 & 4±3 & 18 & 3,896 \\
\cmidrule{2-6}
& Z $\rightarrow$ A $\rightarrow$ B & 0 & 9±8 & 52 & 8,932 \\
\bottomrule
\end{tabular}
\end{table}

\begin{table}[ht]
    \centering
    \renewcommand{\arraystretch}{0.8}
    \caption{Results of IVs Quality and Edge Nodes}
    \label{tab3}
    \begin{tabular}{@{}ccccc@{}}
        \toprule
        \multirow{2}{*}{Dataset} & \multicolumn{4}{c}{Num. of Chain Patterns} \\
        \cmidrule(l){2-5}
        & \multicolumn{1}{c}{\begin{tabular}[c]{@{}c@{}}IV is\\ edge nodes\end{tabular}} & \multicolumn{1}{c}{\begin{tabular}[c]{@{}c@{}}low-\\ quality\end{tabular}} & \multicolumn{1}{c}{\begin{tabular}[c]{@{}c@{}}middle-\\ quality\end{tabular}} & \multicolumn{1}{c}{\begin{tabular}[c]{@{}c@{}}high-\\ quality\end{tabular}} \\
        \midrule
        All-FinCaKG   & 446  & 9,973 & 9,618 & 87 \\
        SH-FinCaKG    & 294  & 3,260 & 5,577 & 95 \\
        ST-FinCaKG    & 345  & 7,811 & 5,701 & 21 \\
        \bottomrule
    \end{tabular}
\end{table}

\begin{table*}
\centering
\small
\caption{Standpoint-based Causality Selection Empirical Analysis Results}
\label{tab4}
\renewcommand{\arraystretch}{0.9}
\begin{tabular}{cllcccc}
\toprule
\textbf{Chain Selection} & \textbf{Node Selection} & \textbf{\begin{tabular}[c]{@{}c@{}}2SLS\\ - 1st /2nd stage\end{tabular}} & \textbf{\begin{tabular}[c]{@{}c@{}}2SLS\\ - coefficient\end{tabular}} & \textbf{t-value} & \textbf{\begin{tabular}[c]{@{}c@{}}Anderson canon.\\ corr. LM statistic\end{tabular}} & \textbf{\begin{tabular}[c]{@{}c@{}}Cragg-Donald(CD)\\ Wald F statistic\end{tabular}} \\
\midrule
\multirow{3}{*}{\textbf{ST-FinCaKG}} & 
Z: economic exposure & \multirow{2}{*}{1st: Z → A} & \multirow{2}{*}{-0.645***} & \multirow{2}{*}{-4.92} & \multirow{3}{*}{24.96***} & \multirow{3}{*}{12.49} \\
& A: ebitda & & & & & \\
& B: corporate governance & 2nd: A → B & 26.100*** & 3.84 & & \\
\midrule
\multirow{3}{*}{\textbf{SH-FinCaKG}} & 
Z: crude oil & \multirow{2}{*}{1st: Z → A} & \multirow{2}{*}{-1.102***} & \multirow{2}{*}{-3.06} & \multirow{3}{*}{11.43***} & \multirow{3}{*}{11.41} \\
& A: operating profit & & & & & \\
& B: marketable securities & 2nd : A → B & 0.340*** & 2.55 & & \\
\bottomrule
\multicolumn{7}{l}{\scriptsize Footnote: Robust t-statistics in parentheses. *** p$<$0.01, ** p$<$0.05, * p$<$0.1} \\
\end{tabular}
\end{table*}

\subsubsection{DFS Result}

Table~\ref{tab2} presents the results of mining IV from FinCaKG using a DFS (Algorithm~\ref{alg1}). ``All-FinCaKG'' is the entire graph containing knowledge from all samples, while ``ST-FinCaKG'' and ``SH-FinCaKG'' are subgraphs that include knowledge from only the ``stakeholder-oriented'' or  ``shareholder-oriented'' standpoints, respectively. Due to the assumption that ``each entity can potentially serve as an IV for other causal pairs ($A \to B$)'', the number of Z equals the total number of entities in the knowledge graph. The table shows that the All-FinCaKG comprises 2,436 IVs, and ST-FinCaKG and SH-FinCaKG subgraphs contain 1,890 and 1,566 IVs, respectively. This indicates that different standpoints may lead to varying causal explanations for business operations.

\begin{figure}[t]
\centerline{\includegraphics[width=0.5\textwidth]{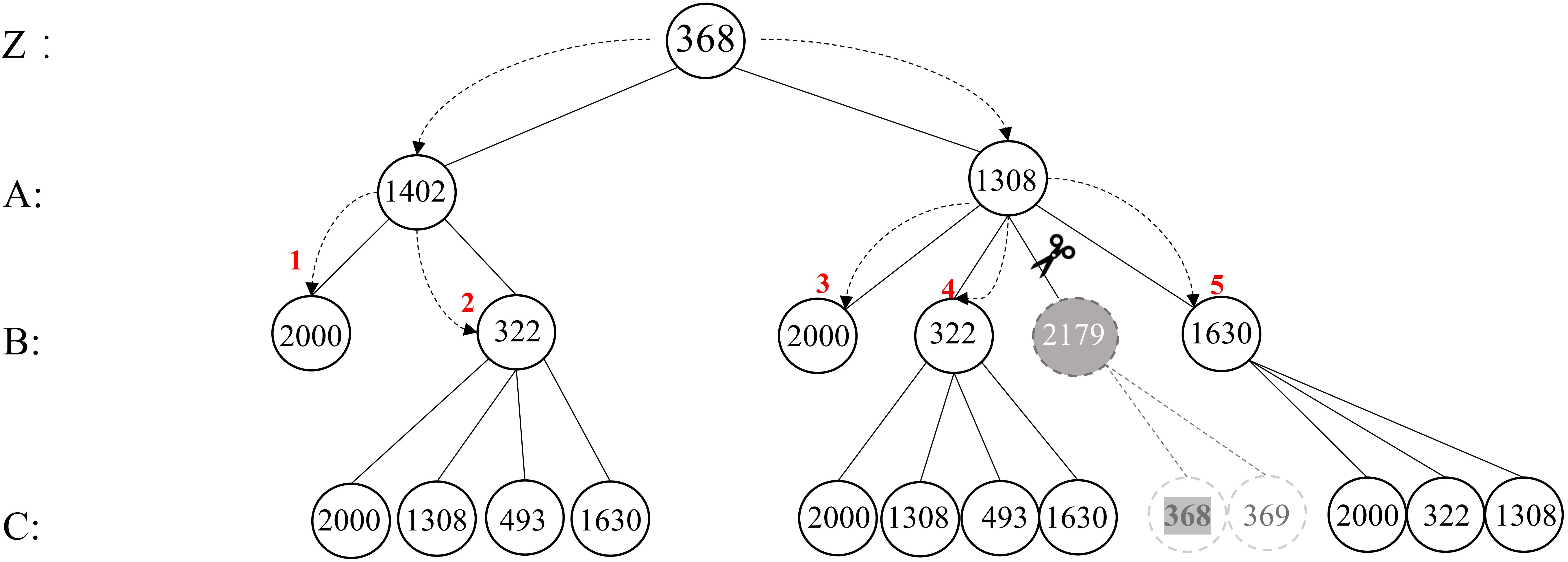}}
\caption{An instance of causal variables identification in DFS algorithm. The dashed arrows indicate a hop in a causal chain. The gray highlight signifies that the node has been removed from the graph (invalid node).}
\label{fig3}
\end{figure} 

$Z \to A \to B$ indicates that Z can only influence B through A. Figure~\ref{fig3} is an instance of DFS, which illustrates the search process when node id ``368'' is selected as an IV. According to the definition, Z is related to A but is not related to B. The association set of 368 consists of 1402 and 1308. Among these, 1402 is related to $\{2000, 322\}$, and 1308 is related to $\{2000, 322, 2179, 1630\}$. However, since the C set related to 2179 contains 368 (Z), this indicates that B is related to Z; consequently, 2179 is removed from the B set. Finally, we can conclude that 368 can serve as an instrumental variable for 5 causal pairs: $\{(1402 \to 2000), (1402 \to 322), (1308 \to 2000), (1308 \to 322), (1308 \to 1630)\}$.

On average, each financial entity can serve as an IV for 13 causal pairs. In the entire knowledge graph, 19,678 patterns meeting the conditions of IV are discovered. In the ST-FinCaKG subgraph, 13,533 qualifying patterns are found; in the SH-FinCaKG subgraph, 8,932 patterns are identified. The number of patterns in the entire graph (19,678) is less than the sum of patterns in the ST-FinCaKG and SH-FinCaKG subgraphs (22,465), indicating some overlaps; i.e., several certain patterns may appear in both subgraphs. This also suggests that each subgraph may have its unique patterns.


\subsubsection{IVs Quality Classification Results}


Table~\ref{tab3} shows the number of low-quality, middle-quality, and high-quality ``\textit{IV-treatment-outcome}'' causality in FinCaKG and subgraphs. SH-FinCaKG identifies the most high-quality IVs and corresponding explainable causal relationships exclusively (95). ST-FinCaKG includes more IVs (1,890) than SH-FinCaKG (1,566) (see Table~\ref{tab2}) However, only 21 causal relationships can be identified using high-quality IVs. Despite ST-FinCaKG capturing more expertise, many of its causal relationships are confounded. SH-FinCaKG provides a clearer and more persuasive explanation for the ``shareholder value maximization'' standpoint compared to ST-FinCaKG. Here is another potential explanation: though ST-FinCaKG has more edge nodes (345), the connections between the variables are not as close as those in SH-FinCaKG (weights are too small.) In other words, the expertise in SH-FinCaKG might be more standpoint-concentrated.

We also compare the IV patterns identified from different subgraphs. There are 870 IVs that appear exclusively in the ST-FinCaKG, 546 specific IVs in the SH-FinCaKG, and 1020 IVs included in both subgraphs. It indicates that most of the expertise are common sense in FincaKG. The question arises: Are the causal relationships observed exclusively in certain subgraphs attributable to standpoint bias of experts, or do they reflect logical patterns specific to certain types of companies? We further discuss this issue using real financial data in the next section.

\subsubsection{Empirical Study of Standpoint Causality}

We validate two standpoint-based causal relationships in ST-FinCaKG ($economic \to EBITDA \to governance$) and SH-FinCaKG ($oil \to profit \to securities$) using high-quality IVs in the overall sample. Table~\ref{tab4} presents the results of 2SLS regression model with industry-fixed effects and year-fixed effects. we represent `\textit{economic exposure}' and `\textit{crude oil}' with foreign exchange exposure and oil price exposure, which are calculated based on the industry's sensitivity to foreign exchange and oil price fluctuations. The ratio of independent directors is used as a proxy variable for corporate governance. This result demonstrates that these specific causal relationships are also significant in the overall sample. The foreign exchange exposure clarifies the positive effect of EBITDA on corporate governance with a coefficient of 26.1 (p$<$0.01), and the oil price exposure identifies the positive effect of operating profit on market securities with a coefficient of 0.34 (p$<$0.01). The Anderson canon. statistics are significant(p$<$0.01) and the CD statistics are greater than 10, suggesting that Z as an instrumental variable satisfies the assumptions of relevance $(Cov(Z, A) \neq 0)$ and exogeneity $(Cov(Z,\varepsilon_2) = 0)$. In this example, the causal relationships observed exclusively in the subgraphs are more likely to reflect standpoint-based biases since they are significantly validated across the overall real-world sample.

\section{Conclusion and Future Work}

Causal inference is an expanding field with a significant impact on both academic research and industrial applications. Our results show that the expertise-driven model provides optimal expertise for financial causal variables identification and interpretation. Additionally, we separate the causal knowledge graph into two subgraphs according to divergent standpoints. We concentrate on those specific causal relationships that are included exclusively in the subgraph. Different from well-researched consensus expertise, these standpoint-based causal relationships are confounded and insufficiently studied. They have the potential to be generalized, thus showing more research opportunities. We interpret the directional causality of two specific expertise by introducing high-quality IVs and demonstrate the significance and applicability of these causal variables in general cases by running the 2SLS regression model. In future practical applications, \textbf{ETE-FinCa} can assist economists in identifying under-explored financial causal relationships and the IV candidates for validation.

\section{ACKNOWLEDGMENT}
This paper is partially supported by the New Energy and Industrial Technology Development Organization (NEDO).

\vspace{12pt}
\color{red}

\end{document}